\documentclass[aps,prl,fleqn,amsmath,twocolumn,longbibliography,footinbib]{revtex4-2}

\usepackage{graphicx}
\usepackage{notes2bib}
\usepackage{etoolbox}
\pretocmd{\eqref}{Eq.\,}{}{}%
\usepackage{amsmath}
\usepackage[charter]{mathdesign}
\DeclareMathAlphabet{\pazocal}{OMS}{zplm}{m}{n}
\def\cal#1{\pazocal{#1}}
\def\d{{\mathrm d}}
\def\tint{\mbox{\large$\int$}}
\def\cbone{\mbox{${\cal C}_{1}$}}
\def\cbtwo{\mbox{${\cal C}_{2}$}}
\def\vbone{\mbox{${\cal V}_{1}$}}
\def\vbtwo{\mbox{${\cal V}_{2}$}}
\def\cb{{\cal C}}
\def\vb{{\cal V}}
\def\b{{\cal B}}

\begin{document}

\title{High harmonics from backscattering of delocalized electrons}
\author{Chuan Yu}
\author{Ulf Saalmann} 
\author{Jan M. Rost}
\affiliation{Max-Planck-Institut f{\"u}r Physik komplexer Systeme,
 N{\"o}thnitzer Str.\ 38, 01187 Dresden, Germany}
\date{\today}

\begin{abstract}\noindent
It is shown that electron backscattering can enhance high-harmonic generation in periodic systems with broken translational symmetry. Paradigmatically, we derive for a finite chain of atoms the harmonic cutoff due to electrons backscattered from the edges of the chain and demonstrate a maximum in the harmonic yield if twice the quiver amplitude of the driven electrons equals the chain length. For an intuitive understanding of our quantum results we develop a refined semiclassical trajectory model with finite electron-hole separation after tunneling. We demonstrate that the same ``tunnel exit'' also holds for interband harmonics in conventional periodic solid-state systems. 
\end{abstract}

\maketitle

\noindent
Since the pioneering experiment by Ghimire et al.\ \cite{ghdi+11} high-harmonic generation (HHG) with strong laser fields applied to solids has been a focus of experimental and theoretical research with first reviews available \cite{vabr17,ghre19,yuji+19}.
The so-called ``three-step model'' \cite{leba+94} is key to understand the microscopic electron dynamics of HHG in atoms and molecules semi-classically in terms of classical trajectories \cite{saro99}. It has been adapted successfully for interband HHG in solids \cite{vamc+14,vamc+15,paer+20}, suggesting that fundamental properties of high harmonics are ruled by the same basic principles from atoms to solids.
On the other hand, a solid-state environment should offer more possibilities to influence these phenomena than an atom due to the larger structural complexity and variability \cite{ndgh+16,tamu+17,baha18,luwo18,siji+19}. Indeed, under suitable conditions, a solid-state HHG spectrum exhibits several cutoffs \cite{ndgh+16,iksh+17} due to the \mbox{(band-)}\discretionary{}{}{}structured continuum of electrons, in contrast to the single atomic cutoff.

In an atomic context, cutoffs can be extended if the laser-driven electron acquires a larger momentum through backscattering from another atom or ion.
This requires a large distance of the order of the atomic quiver amplitude $A_{0}/\omega_{0}$ between the backscattering and recombining ion, where $A_{0}$ is the peak vector potential and $\omega_{0}$ the carrier frequency of the laser. This can theoretically be achieved in laser-assisted ion-atom collisions with a suitable impact parameter \cite{lero03} or for above-threshold ionization in rare-gas clusters with a suitable size, as demonstrated recently in an experiment \cite{waca+20}, but not in molecules which are typically too small.
Solid-like systems, on the other hand, can easily match the spatial requirements set by the quiver amplitude of conduction-band electrons and any irregularity in their periodicity may give rise to backscattering. Indeed, we will analytically predict and numerically demonstrate in the following significantly extended HHG cutoffs through backscattering.

For delocalized electrons, this is to our knowledge a new mechanism which has not been described before. Yet, as in the familiar case of elastic backscattering of localized electrons from a nucleus in atoms or molecules, it is characterized by a reversal of the electron momentum $k\,{\to}\,{-}k$. To be specific, we study backscattering for a finite chain of regularly placed atoms, where the global potential causes backscattering of the delocalized electron wavepacket near the end of the chain with the driving laser field polarized along the chain.
This phenomenon should not be confused with backscattering of localized electrons between two layers of a bi-layer material with laser polarization perpendicular to the layer planes \cite{yuji+20}.

We will show that extended cutoffs through backscattering from the edge can occur and that HHG is most efficient if the full excursion of the excited laser-driven electron (twice the quiver amplitude $x_{\mathrm q}$) matches the length of the chain, i.\,e., if the number of atoms $N\,{\approx}\,N_{\mathrm q}$ with the latter defined by $2 x_{\mathrm q}\,{\equiv}\,N_{\mathrm q} d$ and $d$ denoting the interatomic distance.
Motivated by simple scaling arguments and (semi-)classical trajectory picture for interband harmonics, the predicted cutoff and maximal high-harmonic yield is accurately reflected in the HHG spectra obtained from the laser-driven current.
Atomic units (a.u.) are used throughout unless stated otherwise.

\begin{figure*}[t]
\includegraphics[width=\textwidth]{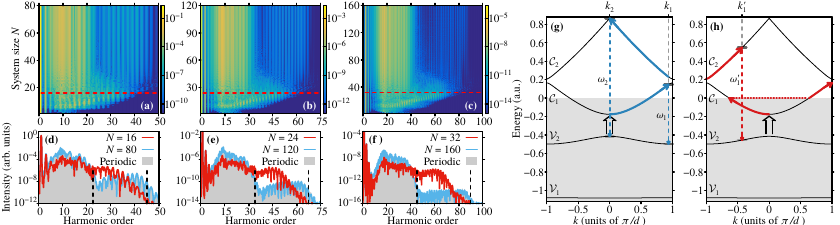}
\caption{HHG spectra as a function of harmonic order and system size $N$ for wavelengths of 1600\,nm (a), 2400\,nm (b) and 3200\,nm (c) and for the same wavelengths but at fixed $N\,{=}\,16$ (d), $N\,{=}\,24$ (e), and $N\,{=}\,32$ (f), indicated by horizontal dashed lines in (a--c). The spectra of the periodic system are shown as shaded area for comparison.
The dashed lines indicate the estimates of the 1st and 2nd cutoff for the periodic system at $\omega_{1}$ and $\omega_{2}$, specified in the text.
(g) Sketch of the $k$-space dynamics in the periodic system, for an electron excited to the bottom of {\cbone} at $A(t_{0})\,{\approx}\,{-}A_{0}$. The maximal {\cbone--\vbtwo} and {\cbtwo--\vbtwo} band energy differences (i.\,e., the cutoffs $\omega_{1}$ and $\omega_{2}$) are achieved at $k_{1}\,{=}\,{2A_{0}}$ and $k_{2}\,{=}\,0$.
(h) Sketch of the $k$-space dynamics in the finite system, with an edge backscattering event in {\cbone} at the vacuum level occurring at $A(t_{\mathrm s})\,{=}\,{-}A_{0}$. The horizontal dotted line represents the sign change of $k(t)$ due to backscattering. With a subsequent band-gap transition to {\cbtwo}, this sketch corresponds to the maximally achievable harmonic energy in the backscattering case $\omega'_{1}$ at $k'_{1}$ (see text).
}
\label{fig:overview}
\end{figure*}%

To keep the situation as simple as possible, we investigate HHG from a chain of $N$ atoms with a lattice constant (interatomic distance) of $d=7$\,a.u.\ and 4 active electrons per atom, as introduced before \cite{hade+17}.
Apart from small modifications we find the electron dynamics in a chain with $N\,{\gtrsim}\,10$ well described with the band structure of the periodic system \bibnote[suppl]{See supplemental material at [url]}.
This is consistent with earlier work \cite{haba+18}, in which the HHG response from a finite chain was found to deviate from that of single atoms or small molecules, revealing a solid-like behavior at rather small system sizes.

For the chain of $N$ atoms we compute the harmonic spectrum generated per atom 
\begin{equation}\label{eq:hhgs_N}
S_{N}(\omega)\propto N^{-2}\Big|\tint\!\d{t}\ J_{\text{tot}}(t)\,W(t)\,\exp({-{\mathrm i}\omega t})\Big|^{2},
\end{equation}
where $J_{\text{tot}}(t)$ is the total current in the system 
and $W(t)$ is a window function of the laser-pulse-envelope shape for improving the signal-to-noise ratio. 
Details of the methods and parameters used as well as the periodic treatment for the limit $N{\to}\infty$ can be found elsewhere \cite{yuir+20}.
The finite chains are treated with density functional theory (DFT) on a real-space grid much larger than the system extension $Nd$ without using periodic boundary conditions; in this way we construct effective (multi-well) potentials with edges self-consistently, and also account for the escape of laser-driven electrons from the system as in a realistic situation \cite{hade+17}.
The laser pulse with frequency $\omega_{0}$ is described in dipole approximation by a vector potential 
$A(t)\,{=}\,A_{0}\sin^{2}[{\omega_{0}t}/({2n_{\text{cyc}}})]\sin(\omega_{0}t)$ for $0\,{\leq}\,t\,{\leq}\,{2\pi n_{\text{cyc}}}/{\omega_{0}}$ and $A(t)\,{=}\,0$ otherwise.
All presented results are obtained with $n_{\text{cyc}}\,{=}\,9$ and $A_{0}\,{=}\,0.21$, but backscattering is not restricted to specific laser parameters as will become clear.

An overview of the results is presented in Fig.\,\ref{fig:overview} for the three laser wavelengths $\lambda \,{=}\,$1600, 2400 and 3200\,nm.
The vertical structure in panels a--c for large $N$ signals that the HHG spectra approach the periodic limit $N{\to}\infty$. However, in the lower central part of Figs.\,\ref{fig:overview}a--c one sees a stronger HHG response which prevails for a certain range of systems sizes. To see this more clearly, Figs.\,\ref{fig:overview}d--f show (in red) spectra at the system sizes $N\,{=}\,16$, 24 and 32, where the HHG response for mid-size harmonic orders is enhanced. 
These selected system sizes (marked by horizontal dashed lines in Figs.\,\ref{fig:overview}a-c) have the widest enhancement region.
The enhancement is particularly evident in comparison to the periodic limit (grey areas). The latter is apparently reached in the longest chains considered for each wavelength ($N\,{=}\,80$, 120 and 160, respectively), which are presented in blue.
All the spectra in Figs.\,\ref{fig:overview}d-f exhibit a sharp rise of intensity when the harmonic energy goes above the {\cbone--\vbtwo} band gap. This indicates that the harmonics above this energy gap are dominated by interband processes, since such a close link between the emitted photon energy and the band energy difference is a clear signature of interband harmonics. As we will see below, the interband recombination picture indeed provides a good interpretation of the spectral shape as well as the time-frequency profile for the high harmonics.

The periodic spectra exhibit two clear steps corresponding to the end (cutoff) of a 1st and 2nd plateau marked by dashed vertical lines. As one can see from the band structures in Fig.\,\ref{fig:overview}g, these cutoffs represent the maximally possible recombination energies \cite{iksh+17} with the valence band \vbtwo$(k)$: $\omega_{1}=\cbone(k_{1})-\vbtwo(k_{1}) = 0.64$ at $k_{1}\,{=}\,2A_{0}\,{=}\,0.94\,\pi/d$ and $\omega_{2}=\cbtwo(k_{2})-\vbtwo(k_{2}) = 1.28$ at $k_{2}\,{=}\,0$.
The large gaps to all other bands prevent {\vbone} to actively participate in the HHG processes, as discussed before \cite{hade+17}.
An electron, excited from the 2nd valence band {\vbtwo} at $t_{0}$, preferentially enters the 1st conduction band {\cbone} at $k_0\approx 0$ near the $\Gamma$ point ($k\,{=}\,0$) and subsequently moves with momentum 
\begin{equation}\label{eq:normal_k}
k(t)=A(t)-A(t_{0})+k_{0}\,.
\end{equation}
This time-dependent $k$-space motion always holds in the periodic limit, but can be modified by backscattering in finite systems, as will be discussed below.

The enhanced spectra (red in Fig.\,\ref{fig:overview}d--f) exhibit a small dip at the 1st cutoff but the enhancement does not extend to the 2nd cutoff. This observation suggests that the enhancement is not due to a more efficient mechanism to enter {\cbtwo} preserving the original $k(t)$. Rather, it must be a process which changes $k(t)$. This can be achieved by elastic scattering in the presence of a laser field. Indeed, as we will see, the enhancement is due to electrons in {\cbone}, being backscattered from the edge of the chain. To this end, we have to understand how elastic scattering in real space manifests itself in the band picture of reciprocal space.

When an electron wavepacket approaches the system edge, it can either be reflected from it (i.\,e., being backscattered) or leak out of the system (i.\,e., being ionized). Backscattering|ionization will be dominant if its mean energy is lower|higher than the vacuum level.
In the classical three-step description, backscattering of a laser-driven localized electron at a time $t_{\mathrm s}$ is assumed to be elastic, resulting in a sign change of the electron's instantaneous momentum.
In a quasi-periodic system, the delocalized electron (and the accompanying hole) suffer the analogous momentum kick while moving on their respective band $\b$ with dispersion $\b(k)$, i.\,e.,
\begin{equation}\label{eq:backscat_k}
k(t\,{>}\,t_{\mathrm s})=A(t)-2A(t_{\mathrm s})+A(t_0)-k_{0}.
\end{equation}
This is illustrated in Fig.\,\ref{fig:overview}h for the electron.
Note that the reversal of the momentum indicated by the horizontal dotted line is an essential signature of elastic backscattering that is distinct from normal intraband motions such as dynamical Bloch oscillations \cite{scho+14} in which $k(t)$ does not jump.
In general, the band energy at backscattering must be below the vacuum level $E\,{=}\,0$ to avoid ionization. Therefore, the maximal momentum at backscattering is $k_{\mathrm s}\,{=}\,0.285$, defined by $\cbone(k_{\mathrm s})\,{=}\,0$. Since the electron can acquire at most a momentum of $2A_{0}$ through (unperturbed) interaction with the laser field, the maximal final momentum is $k'_{1}\,{=}\,k_{\mathrm s}{+}2A_{0}{-}2\pi/d$ in the first Brillouin zone (BZ) leading to the recombination energy $\omega'_{1}\,{=}\,\cbtwo(k'_{1})-\vbtwo(k'_{1})\,{=}\,1.0$ which defines the 1st cutoff energy \emph{extended through backscattering}.
Indeed, this corresponds to harmonic order $35,53,70$ for the wavelengths $\lambda=1600,2400,3200$\,nm, respectively, where the yield of the enhanced spectra (red) in Figs.\,\ref{fig:overview}d--f decreases.

In passing we note, that without backscattering, the electron in Fig.\,\ref{fig:overview}h would never have the chance to pass the BZ boundary and enter {\cbtwo} via a subsequent band-gap transition.
Hence, edge backscattering suggests itself as a pathway to high-energy states in analogy to backscattered electrons from an ion in the atomic context. There, however, backscattering only leads to higher photo-electron energies \cite{pabe+94,waca+20}, but not to larger cutoffs in HHG.
This is mainly due to the fact that the electron's wavefunction in the atomic context is usually spatially localized on the ion (playing the role of the hole) and the continuum electron wavepacket. The lacking overlap prevents recombination necessary for HHG between the energetic electron far away from the ion available for recombination.
In solid-like systems, on the other hand, we deal with spatially delocalized Bloch electrons, for which overlap of electron-hole wavefunctions can be more easily achieved \cite{yuga20}.
Moreover, delocalized electrons reflected by the edges continue to move inside the system, allowing them to recombine with significant wavefunction overlap.
Therefore, backscattering represents a promising mechanism for increasing the energy of solid-state harmonics.

\begin{figure}[t]
\includegraphics[width=\columnwidth]{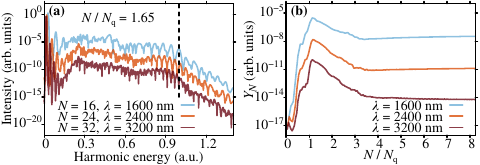}
\caption{The HHG spectra from Fig.\,\ref{fig:overview}d--f as function of photon energy (a) and their integrated yield $Y_N$ beyond the 1st cutoff $\omega_{1}$ as a function of scaled system size $N/N_{\mathrm q}$ (b). The vertical dashed line in panel (a) represents the edge-backscattering cutoff at $\omega'_{1}\,{=}\,1.0$, discussed in the text.
}
\label{fig:comparison}
\end{figure}

As a next step we work out which role the spatial extension of the chain plays for backscattering. To this end we vary in Fig.\,\ref{fig:comparison} the wavelength $\lambda$ while keeping the vector potential amplitude $A_{0}$ fixed.
The latter ensures that the dynamics in momentum space, and in particular the energy gain through backscattering depending on $A_{0}$ as discussed so far, remains the same while through the variation of the wavelength the quiver amplitude $x_{\mathrm q}\propto A_{0}\lambda$ changes linearly, resulting in different scales for the spatial dynamics.
Hence, locking the ratio of chain length versus wavelength $N/\lambda$ in addition to an identical $A_{0}$ should provide similar conditions for the high-harmonics-generating electron dynamics and we expect similar spectra, provided the HHG yield is shown as a function of harmonic energy, as done in Fig.\,\ref{fig:comparison}a.
The similarity of the three spectra with the extended cutoff at $\omega'_{1}\,{=}\,1.0$ is evident.

In Fig.\,\ref{fig:comparison}b we demonstrate that this similarity implies a universal ratio $N_{\mathrm q}/\lambda$ where the largest enhancement of the high-harmonic yield for edge backscattering occurs. For this purpose we integrate the yield in the spectral region of enhancement, $Y_{N}=\int_{\omega_{1}}\!\!\d{\omega}\ S_{N}(\omega)$.
That the curves level off for large $N$ simply reflects convergence to the periodic limit without edge backscattering. That all three integrated yields have a similar shape over the entire scaled range of $N$ illustrates the universality of the underlying strong-field dynamics of delocalized electrons provided that momentum and spatial dynamics is equivalent.
Most interesting in the context of backscattering is the sharp rise and maximum of $Y_{N}$ which occurs close to $N/N_{\mathrm q}\,{=}\,1$ for different wavelengths with the critical number of atoms $N_{\mathrm q}=2x_{\mathrm q}/d$, where the length $N_{\mathrm q}d$ of the chain equals the full quiver excursion $2 x_{\mathrm q}$ of the excited electron.
This 
can be understood by considering the extreme cases: For $2 x_{\mathrm q}\,{\ll}\,N d$ only a a small fraction of excited electron density can reach the chain edge for backscattering. The limit $2 x_{\mathrm q}\,{\gg}\,N d$ implies that $A(t_{\mathrm s})$ and $A(t_0)$ hardly differ and therefore the momentum gain at backscattering $\Delta k\,{=}\,-2k(t_{\mathrm s})\,{=}\,2[A(t_0)-A(t_{\mathrm s})]$, with the band-gap tunneling assumption $k_0\,{=}\,0$, becomes negligible.
Hence, $N d\,{=}\,2 x_{\mathrm q}$ is the optimal length where the entire spatially distributed excited electron density can participate in backscattering.

We note that while the momentum scale $A_{0}$ is a property of the light only, this is not the case for the spatial scale $x_{\mathrm q}$, the quiver amplitude, which depends also on the band structure.
With $k(t)$ given by \eqref{eq:normal_k}, the position-space motion of a Bloch electron in band $\b$ reads
\begin{equation}\label{eq:position_x}
\!\!\!\!\!\!
\Delta x_{\b}(t) \equiv x_{\b}(t)-x_{\b}(t_{0}) = \tint_{\!\!t_0}^{t}\d{t'}\mbox{$\frac{\d}{\d{k}}$}\b(k)\big|_{k=k(t')}.
\end{equation}
Within the Kane band approximation \cite{ka57}, an explicit expression for the quiver motion can be given which is even analytically solvable if the electron moves with initial condition $A(t_{0})\,{=}\,0$ in the conduction band \citenote{suppl}.
In this case $x_{\mathrm q}= [A_{0}/(m_{*}\omega_{0})] \arctan(a)/a$, with $a\,{=}\,A_{0}/k_{*}$, where $m_{*}$ is the effective mass of the electron and $k_{*}$ the band's momentum scale 
\bibnote{For our system the parameters $m_{*}\,{=}\,0.167$ and $k_{*}\,{=}\,0.2$ lead to $x_{\mathrm q}\,{=}\, 4.62\,A_{0}/\omega_{0}$ and $N_{\mathrm q}\,{=}\, \lambda/165\,\text{nm}$.}.

\begin{figure}[t]
\includegraphics[width=\columnwidth]{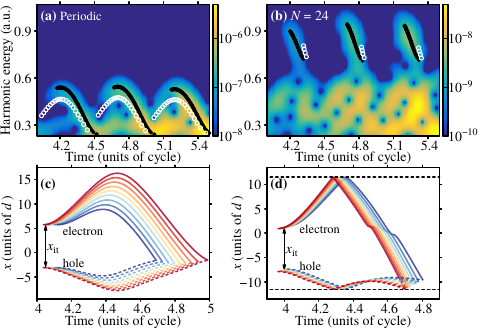}
\caption{HHG time-frequency profile at $\lambda\,{=}\,2400$\,nm for the periodic system (a) and a finite chain of $N\,{=}\,24$ (b), respectively.
The black dots are traces obtained from our refined semiclassical recollision trajectories with finite electron-hole separation $x_{\mathrm{it}}$ after tunneling.
Consistent with the Taylor expansion (see text) we use the same $x_{\mathrm{it}}\,{=}\,{-}E_{\mathrm g}/F(t_{\mathrm m})$ for all trajectories in the small interval $|t_0{-}t_{\mathrm m}|<0.25\pi/\omega_0$ around $t_{\mathrm m} = 3.5T,4.0T,4.5T$ contributing to the traces. 
The white circles are traces obtained from trajectories assuming $x_{\mathrm{it}}\,{=}\,0$ without complex initial time, see text.
Panels (c) and (d) provide representative trajectories forming the black traces in (a) and (b) in real space.
The horizontal dashed lines in (d) indicate the locations of the edge atoms, where backscattering is assumed to occur.
}
\label{fig:mechanism}
\end{figure}%

Finally, we discuss how the HHG time-frequency profile, shown in Fig.\,\ref{fig:mechanism}, obtained by Gabor transforming the quantum current in \eqref{eq:hhgs_N}, can be mapped onto classical trajectories from \eqref{eq:position_x} for electron-hole pairs.
It is \mbox{a priori} unclear if the (quantum) reflection-based backscattering mechanism can be described adequately with trajectories. Certainly, such trajectories will require refined spatial properties. To identify them, we first analyze the (standard) periodic system case.

To fulfill the stationary-phase condition for the tunneling step $E_{\cb\vb}[k(t_{\mathrm i})]\equiv\cbone[k(t_{\mathrm i})]-\vbtwo[k(t_{\mathrm i})] =0$ justifying the trajectory picture in the first place \cite{vabr17}, we propagate electron and hole trajectories from an initial complex time $t_{\mathrm i}\,{=}\,t_0\,{+}\,{\mathrm i}\tau$ over an imaginary time span ${\mathrm i}\tau$ realizing the tunneling process.
Tunneling happens most likely near the band gap $E_{\mathrm g}$ at the $\Gamma$ point ($k\,{=}\,0$), where each band typically has an approximately parabolic dispersion around its local extremum.
With an effective mass, we can write for small $|k|$ the difference in the band dispersion as
$E_{\cb\vb}(k) \approx E_{\mathrm g}\,{+}\,{k^2}/({2\mu})$ with $\mu^{-1} \equiv \tfrac{\d^2}{\d{k}^2}{E}_{\cb\vb}(k)\big|_{k=0}$.
Hereby $\mu$ is the \emph{reduced} effective mass of the electron-hole pair, which has for the system considered here the value $\mu\,{=}\,0.108$.

Then we approximately solve ${E}_{\cb\vb}[k(t_0{+}{\mathrm i}\tau)] = 0$ describing tunneling with complex time and momentum.
The trajectories for interband HHG are typically created when the laser field $F(t)\,{=}\,{-}\mbox{$\frac{\d}{\d{t}}$}A(t)$ is near its extrema at times $t_{\mathrm m}$ fulfilling $\mbox{$\frac{\d}{\d{t}}$}F(t)\big|_{t=t_{\mathrm m}}\,{=}\,0$.
This leads to a relatively small $\tau$, for which we make a (truncated) Taylor expansion $A(t_{ 0}{+}{\mathrm i}\tau) \approx A(t_{ 0}) - {\mathrm i}\tau F(t_{ 0})$ and therefore $k(t_{ 0}{+}{\mathrm i}\tau) \approx -{\mathrm i}\tau F(t_{ 0})$.
Hence, $\tau$ is approximately given by $\tau^2\,{=}\,2 \mu E_{\mathrm g} / F^2(t_{ 0})$.
Denoting the electron-hole separation by $\Delta_{\cb\vb}(t) \equiv x_{\cb}(t)-x_{\vb}(t)$, and integrating the trajectory from $t_0{+}{\mathrm i}\tau$ to $t_0$ along the imaginary-time axis leads to a ``tunnel exit''
\begin{align}\label{eq:tunnel}
x_{\mathrm{it}} & \equiv  \Delta_{\cb\vb}(t_0) - \Delta_{\cb\vb}(t_0+{\mathrm i}\tau) \nonumber\\
& = \tint_{\!\!t_0+{\mathrm i}\tau}^{t_0}\d{t'}\tfrac{\d}{\d{k}}{E}_{\cb\vb}[k(t')] 
\approx -{F(t_0)\tau^2}/2 \mu \nonumber\\
& = - E_{\mathrm g} / F(t_0)
\approx -E_{\mathrm g} / F(t_{\mathrm m}),
\end{align}
which defines the separation of electron and hole trajectories when they start their dynamics at real time $t_0\,(\,{\approx}\,t_\mathrm{m})$ on the conduction and valence band, respectively.

Before tunneling at the complex time $t_0{+}{\mathrm i}\tau$, electron and hole are at the same position, i.\,e., $\Delta_{\cb\vb}(t_0{+}{\mathrm i}\tau) = 0$. 
After tunneling in imaginary time, however, when the trajectory starts propagating in real time at $t_0$, the electron-hole separation is $\Delta_{\cb\vb}(t_0) = x_{\mathrm{it}}$, approximated in \eqref{eq:tunnel}.
Accordingly, the recombination condition at time $t_{\mathrm r}$ is $\Delta_{\cb\vb}(t_{\mathrm r}) = \Delta_{\cb\vb}(t_0+{\mathrm i}\tau) = \Delta_{\cb\vb}(t_0) - x_{\mathrm{it}} = 0$,
which naturally defines the electron-hole recollisions as harmonic emission events.
Note that in the standard solid-state trajectory model $x_{\mathrm{it}}\,{=}\,0$ is assumed \cite{vabr17,vamc+15}.
In Fig.\,\ref{fig:mechanism}a one sees that the trajectories starting with $x_{\mathrm{it}}\,{=}\,{-}E_{\mathrm g}/F(t_{\mathrm m})$, shown in black, track the HHG profile much better than the ones with $x_{\mathrm{it}}\,{=}\,0$, shown in white, which have been used so far.

In the case of backscattering for a finite chain, trajectories with the same tunnel exit $x_{\mathrm{it}}$ propagate until the electron-hole separation reaches the chain length and if the energy of the electron is below the vacuum level, namely $\cbone(t_{\mathrm s})\,{<}\,0$, backscattering takes place by elastic reflection of the trajectories at the chain edges (horizontal dashed lines in Fig.\,\ref{fig:mechanism}d). This means that for $t\,{>}\,t_{\mathrm s}$ \eqref{eq:backscat_k} holds instead of \eqref{eq:normal_k}. These trajectories (black in Fig.\,\ref{fig:mechanism}b) trace the quite different HHG profile very well, while trajectories with $x_{\mathrm{it}}\,{=}\,0$ (white) disagree with the quantum profile.
We may conclude that semiclassical trajectories that include initial propagation in imaginary time lead to a finite electron-hole separation $x_{\mathrm{it}}$ after tunneling, which should be taken instead of $x_{\mathrm{it}}\,{=}\,0$ for condensed-matter interband harmonics in a fully periodic system as well as for the new harmonics from backscattering in a finite chain.

To summarize, we have established backscattering of delocalized electrons as a mechanism to extend the cutoff for harmonics in quasi-periodic systems with an inherent length scale due to broken translational symmetry.
For simplicity and consistency, we have chosen to demonstrate and analyze backscattering with finite chains of atoms solving the many-electron dynamics based on DFT.
This has allowed us to link the quiver amplitude of the driven electron to the extension of the system, revealing that one achieves the highest integrated harmonic yield beyond the 1st cutoff of the fully-periodic system, if twice the quiver amplitude is approximately equal to the length of the atomic chain. The band energy at the momentum where backscattering takes place must be below the vacuum level of the system, otherwise ionization outweighs reflection. This is a universal condition for the extended cutoff, which takes, however different values depending on the band structure.

High harmonics due to backscattering can be described in terms of a simple trajectory picture with elastic reflection from the edges of the atomic chain and a finite initial separation $x_{\mathrm{it}}$ for the electron-hole pair determined by tunneling from valence to conduction band in imaginary time. We have shown that the same tunnel exit $x_{\mathrm{it}}$ also governs interband harmonics in a conventional periodic system, improving the agreement of the trajectory traces with the quantum energy-time profile of the harmonics.

Backscattering as introduced here has close analogies in extended atomic systems. However, in the latter it leads only to higher energies in laser-driven photo-ionization (often termed above-threshold ionization), but not to larger high-harmonic cutoffs, since the localized electrons in atomic systems lack the ability for overlap of electron amplitudes at large distances which is possible for the delocalized electrons in quasi-periodic systems.
Other sources of breaking the periodicity of solid-state systems, such as impurities, domain walls or grain boundaries, may also induce backscattering and ensuing effects on HHG.
Work in this direction is underway.

\begin{acknowledgments}
CY acknowledges discussion with Lars Bojer Madsen in the early stage of this work.
\end{acknowledgments}

\vfill
\def\articletitle#1{\textit{#1}.}

\end{document}